Title (Intelligent breathing soliton generation in ultrafast fibre lasers)

*Xiuqi Wu[1], Junsong Peng[1,2],*, Sonia Boscolo[3], Ying Zhang[1], Christophe Finot[4], and Heping Zeng[1,2],* **

*Corresponding Author: jspeng@lps.ecnu.edu.cn; hpzeng@phy.ecnu.edu.cn

[1]State Key Laboratory of Precision Spectroscopy, East China Normal University, Shanghai 200062, China

[2]Chongqing Institute of East China Normal University, Chongqing 401120, China

[3]Aston Institute of Photonic Technologies, Aston University, Aston Triangle, Birmingham B4 7ET, UK

[4]Laboratoire Interdisciplinaire Carnot de Bourgogne, UMR 6303 CNRS – Université de Bourgogne Franche-Comté, F-21078 Dijon Cedex, France

Abstract: Harnessing pulse generation from an ultrafast laser is a challenging task as reaching a specific mode-locked regime generally involves adjusting multiple control parameters, in connection with a wide range of accessible pulse dynamics. Machine-learning tools have recently shown promising for the design of smart lasers that can tune themselves to desired operating states. Yet, machine-learning algorithms are mainly designed to target regimes of parameter-invariant, stationary pulse generation, while the intelligent excitation of evolving pulse patterns in a laser remains largely unexplored. Breathing solitons exhibiting periodic oscillatory behavior, emerging as ubiquitous mode-locked regime of ultrafast fibre lasers, are attracting considerable interest by virtue of their connection with a range of important nonlinear dynamics, such as exceptional points, and the Fermi-Pasta-Ulam paradox. Here, we implement an evolutionary algorithm for the self-optimisation of the breather regime in a fibre laser mode-locked through a four-parameter nonlinear polarisation evolution. Depending on the specifications of the merit function used for the optimisation procedure, various breathing-soliton states are obtained, including single breathers with controllable oscillation period and breathing ratio, and breather molecular complexes with a controllable number of elementary constituents. Our work opens up a novel avenue for exploration and optimisation of complex dynamics in nonlinear systems.

## 1. Introduction

Breathing solitons form an important part of many different classes of nonlinear wave systems, manifesting themselves as localised temporal/spatial structures that exhibit periodic oscillatory behavior. Breathers were first studied experimentally in Kerr fibre cavities[1] and subsequently reported in optical micro-resonators,[2-4] where the dynamics are governed by the Lugiato-Lefever equation. Recently, they have also emerged as an ubiquitous mode-locked regime of ultrafast fibre lasers.[5-7] These nonlinear waves are attracting significant research interest in optics by virtue of their strong connection with the Fermi-Pasta-Ulam recurrence[8, 9]– a paradoxical evolution of nonlinearly coupled oscillators that periodically return to their initial state, the formation of rogue waves,[10, 11] turbulence, and modulation-instability phenomena. Breathers are also closely related to exceptional points,[6] thereby opening a novel avenue for the exploration of exceptional-point physics. Besides their significance in nonlinear science, breathers also bear interesting possibilities for practical applications. For instance, they can increase the resolution of dual-comb spectroscopy,[12] and the breather regime in a laser oscillator can be used to generate high-amplitude ultrashort pulses without additional compressors.[13, 14]

In addition to their growing use as sources of ultrashort pulses for many applications, mode-locked fibre lasers constitute an ideal platform for the fundamental exploration of complex nonlinear wave dynamics. Indeed, the high levels of linear and nonlinear effects accumulated during a single round trip (RT) in such lasers, along with the nonlinear polarisation evolution (NPE) effect - which is often exploited as the saturable absorption mechanism to drive the mode-locking process[15] - entail a wealth of complex short-pulse dynamics[16] that can be accessed through tuning of the control cavity parameters. These include pulsating regimes,[17] soliton explosions,[17-20] and various multi-pulsing regimes, where the interactions between the solitons lead to harmonic mode locking or self-organised patterns such as soliton bunches and molecules,[21-24] soliton molecular complexes[25] and supramolecular structures.[26] Chaotic pulsed states also abound, including noise-like pulse emission[27] and rogue-wave formation.[28] Operating regimes in which a laser oscillator generates strongly breathing solitons were predicted theoretically in Ref [13]. In Ref [5], we have directly resolved the fast temporal and spectral dynamics of breathers in a mode-locked fibre laser using real-time detection techniques. More recently, we have also reported on the observation of different types of breather molecular complexes (BMCs) in a fibre laser, whose formation is driven by the dispersive waves that are emitted by breathers in the anomalous-dispersion propagation regime.[29] However, reaching a desired operating regime in a fibre laser generally depends on precisely adjusting multiple parameters in a high-dimensional space, which is usually performed through an often lengthy, trial-and-error experimental procedure, due to the lack of analytic relationship between the cavity parameters and the pulse features. The practical difficulties associated with such a procedure – including the problem of repeatability –, along with the inadequacy of systematic numerical propagation modelling, hamper the possibilities of obtaining an augmented subspace of useful ultrafast dynamics through intracavity parameter adjustment.

Nevertheless, such a difficulty can be circumvented by machine-learning strategies and the use of evolutionary and genetic algorithms, which are well-suited to the global optimisation problem of complex functions.[30, 31] The application of advanced algorithmic tools and adaptive feedback and control has recently greatly boosted the progress in the search for a truly self-optimising laser, and a number of groups have reported on different approaches to automate optimisation of one or more parameters of the laser cavity to reach and maintain a desired operating state.[32-41] A recent work[42] introduced extra novelty by incorporating fast spectral measurements into the feedback loop of the laser setting which, along with an intelligent polarisation search algorithm, enabled real-time control of the spectral width and shape of ultrashort mode-locked pulses. Despite these significant advances, the intelligent generation of breathing solitons in a fibre laser remains challenging because breathers refer to a highly dynamical state in which the pulse spectral and temporal characteristics change drastically within a period of oscillation, while existing machine-learning strategies are mostly designed to target laser generation regimes of parameter-invariant, stationary pulses.

In the present experimental work, we implement an evolutionary algorithm (EA) for the self-optimisation of the breather regime in a mode-locked fibre laser, based on the optimal four-parameter tuning of the intracavity nonlinear transfer function through electronically driven polarisation control. We define compound merit functions relying on the characteristic features of the radiofrequency (RF) spectrum of the laser output, which are capable to locate various self-starting breather regimes in the laser, including single breathers with controllable breathing ratio and period, and BMCs with a controllable number of elementary constituents.

## 2. Results

### 2.1. Experimental setups and principle

The experimental setup is sketched in Figure 1(a). The laser is a fibre ring cavity in which a 1.3-m-long erbium-doped fibre constitutes the gain medium, pumped by a laser diode operating at 980 nm through a wavelength-division multiplexer. Other fibres in the cavity are a section of dispersion-compensating fibre and pieces of standard single-mode fibre from the pigtails of the optical components used. The group-velocity dispersion values of the three fibre types are 65, 62.5, and –22.8 $ps^2$/km, respectively, yielding a normal cavity dispersion of 0.028 $ps^2$ at the operating wavelength of ~1.5 μm. The repetition rate of the laser is 16.765 MHz. By using a normally dispersive laser cavity, the generated breathing solitons appear to be more robust than in the anomalous-dispersion case. Indeed, with anomalous dispersion, the Kelly sidebands radiating from the solitons may cause energy variations that do not refer to a breather state, so that the design of an efficient EA for optimisation of the breather laser regime may become more challenging. More importantly, since pulses generated in the anomalous-dispersion regime are typically much shorter, the time resolution of the detection system used may not enable capturing the changes experienced by the temporal duration of breathers within an oscillation period[8]. The mode-locked laser operation is obtained thanks to an effective saturable absorber based on the NPE effect.[15] The nonlinear transfer function of the NPE-based mode locking is controlled by an electronically driven polarisation controller (EPC) working together with a polarisation-dependent isolator. The EPC consists of four fibre squeezers oriented at 45 ° to each other and each controlled by an applied voltage signal and can generate all possible states of polarisation over the Poincaré sphere, where each set of voltages corresponds to a specific state. Therefore, only one EPC is required in the laser setup to achieve the manipulation of the intracavity polarisation states enabling a complete control of the nonlinear transfer function, as in[32] [36, 38, 43]. Furthermore, with a response time of 0.4ms, the EPC can change the laser operation regime quickly, thus enabling the implementation of an EA to find a user-defined optimal laser regime within a realistic time scale. The laser output is split into two ports: a fraction is directly detected by a fast photodiode (Finisar XPDV2320R; 20-ps response time, 50-GHz bandwidth) plugged to a real-time oscilloscope (Agilent; 33-GHz bandwidth, 80-GSa/s sampling rate). The so measured time traces of one dimensional intensity, $I(t)$, in real time are then used to construct the spatio-temporal intensity evolution $I(t,z)$[44] to characterise the time-domain dynamics of the laser. The remaining laser output is sent through a time-stretch dispersive Fourier transform (DFT) setup consisting of a long segment of normally dispersive fibre that provides a total accumulated dispersion of $DL \approx -1200$ ps/nm. After propagation through such a long dispersive link, a short laser pulse gets significantly stretched so that its spectrum is mapped into the time domain.[45] From the photodetection of the DFT output signal on a fast photodiode, the optical spectrum for each pulse is obtained directly on the oscilloscope, with a resolution of $\Delta\lambda = 1/(DL*BW) \approx 0.025$ nm, where $BW$ is the bandwidth of the photodetection. The oscilloscope is connected to a computer that runs the EA and controls the EPC via a field programmable gate array (FPGA) and four digital-to-analogue converters (DACs; 125-MHz sampling rate, 14-bit resolution). During the searching process, the signals generated by the algorithm are delivered to the FPGA, which adjusts the control voltages of the EPC through the DACs. The DACs translate the instructions from the FPGA to the voltages and finally act on the EPC.

The intelligent search of breathers is realised via an EA whose principle, as illustrated in Fig. 1(b), mimics mechanisms inspired by Darwin's theory of evolution: individuals composing a population progress through successive generations only if they are among the fittest.[46] In our case, an individual is a laser regime, associated with the nonlinear transfer function defined by the four control voltages applied to the EPC; these voltages are therefore the genes of the individuals. The process begins with a collection of individuals or 'population' (making up the first generation), each comprising a set of randomly assigned genes. The system output is measured for each individual in the generation, evaluated by a user-defined merit function (also known as fitness or objective function) and assigned a score. The EA then creates the next generation by breeding individuals from the preceding generation, with the probability that an individual is selected to be a 'parent' based on their score ('roulette wheel' selection,[46] Fig. 1(c)). While elitist selection - in which the best individuals are cloned to the next generation to ensure their high-quality genes are preserved -, works well to search for stationary mode-locking states,[40] roulette selection is found more efficient for the breather mode locking here. Two new individuals – children - are created from the crossover of two randomly selected parents, namely the interchange of their genes. A mutation probability is also specified, which can randomly alter the children's genes, thus allowing for the genetic sequence to be refreshed. This process repeats until the algorithm converges and an optimal individual is produced.

The algorithm is initialised with a population of 100 individuals, and the population size of the next generations is kept constant to 50 individuals. This relatively large population size is crucial to targeting the breather mode-locking regime, which exists in a narrower parameter space than stationary mode locking.[14] Crossover is realised by generating random numbers from 0 to 1 and correlating them to each pair of individuals. If a pair has a correlated number smaller than 0.6, crossover within the pair occurs; otherwise, no crossover happens. Mutation is realised in a similar manner, with a smaller probability: an individual is subject to mutation if its correlated number is smaller than 0.02. The mutation implementation is described in the Supporting Information. Evaluation of the properties of an entire generation of individuals typically takes 3.3 minutes.

A critical factor to the success of a self-optimising laser implementation is the merit function, which must return a higher value when the laser is operating closer to the target regime. Merit functions based on the peak height of the cavity repetition frequency or pulse count work well for searching stationary mode locking regimes[38, 42] but cannot be used as standalone to target breather modes of operation. Therefore, we formulate a compound merit function based on the following observations. To select a breather regime, we need a merit function that discriminates between breather and stationary pulsed operations. The oscillation frequency of the breathers manifests itself as sidebands in the RF spectrum of the laser output, as illustrated in Fig. 1(d), where $|f_{\pm 1} - f_r|$ represents the breathing frequency, and $f_r$ and $f_{\pm 1}$ are the cavity repetition frequency and sideband frequencies, respectively. There are no sidebands located at $f_{\pm 1}$ when the laser works in a stationary mode locking regime. Therefore, we can design a merit function that exploits the intensity ratio of the central band located at $f_r$ to the sidebands at $f_{\pm 1}$. In practice, however, it is easier to operate with the intensity ratio of $f_r$ to the frequency interval from $f_{-1}$ to $f_{+1}$. Hence, the merit function for the breather operation derived from this feature is

$$C_b = \frac{\sum_{f=f_-}^{f=f_+} I(f) - \sum_{f=f_r-\Delta}^{f=f_r+\Delta} I(f)}{\sum_{f=f_{-1}}^{f=f_1} I(f)} \quad (1)$$

where $\sum_{f=f_-}^{f=f_+} I(f)$ and $\sum_{f=f_r-\Delta}^{f=f_r+\Delta} I(f)$ are the intensities measured across the frequency interval from $f_-$ to $f_+$ and the width $2\Delta$ of the frequency band centred on $f_r$, respectively. Therefore, the difference between these two quantities relates to the intensities of the sidebands at $f_{\pm 1}$, hence $C_b$ represents the relative strength of the sidebands with respect to the frequency interval [$f_{-1}$, $f_{+1}$]. Accordingly, if $C_b$ approaches zero, it means that $f_r$ prevails and there are no sidebands, indicating a stationary mode locking state. On the contrary, a $C_b$ value far from zero evidences the presence of strong sidebands in the RF spectrum, indicating possible breather formation in the laser cavity. In our experiments, the RF spectrum is obtained directly from the oscilloscope that processes the fast Fourier transform of the laser output intensity recording, and a span of 381.44 KHz ($f_r \pm 190.72$ KHz) is chosen for the measurements.

However, equation (1) alone cannot guarantee the generation of breathers in the laser because other laser modes, such as relaxation oscillations or noise-like pulse emission, may also feature sidebands in the RF spectrum (see Supplementary Figs. 1 and 2). Therefore, to exclude these possibilities, we use the merit function relating to the mode-locked laser operation[38], which is derived from the feature that mode-locked pulses have a significantly higher intensity than free-running states,

$$C_{ml} = \frac{\sum_{i=1}^{i=L} I_i}{L}, I_i = \begin{cases} I_i, (I_i \geq I_{th}) \\ 0, (I_i < I_{th}) \end{cases} \quad (2)$$

where $L$ is the number of laser output intensity points recorded by the oscilloscope ($L=2^{24}$, corresponding to a time trace of ~2700 cavity RTs), $I_i$ is the intensity at point $i$ and $I_{th}$ is a threshold intensity that noise should not exceed. Therefore, $C_{ml}$ represents the average of pulses' intensities. We can then define the total merit function of the breather mode-locking regime as the weighted sum of $C_{ml}$ and $C_b$,

$$F_{merit} = \alpha \times C_{ml} + \beta \times C_b \quad (3)$$

where the weights of the two components are determined empirically ($\alpha=40000$, $\beta=10$).

The search for single-breathing soliton mode locking is implemented here as a three-stage optimisation procedure. The first stage involves testing and ranking each individual according to its fitness with respect to the merit function $C_{ml}$ which enables the exclusion of those individuals that correspond to relaxation oscillation regimes whose merit score is low. Meanwhile, to exclude noise-like pulse mode locking, which returns moderate values of $C_{ml}$ similar to stationary mode locking, the maximum peak intensity of the pulses is checked: if it is extremely high, a new search begins. The second stage involves pulse counting to select mode locking at the fundamental repetition frequency (the details of pulse count are provided in the Supporting Information). Finally, the individuals passing through the first two stages, are scored against the compound merit function given in equation (3) to exclude stationary pulse states. The inclusion of additional components in the definition of $F_{merit}$ is the key to achieving advanced control of the characteristics of the breather state, such as tuning of the oscillation period or breathing ratio. The corresponding merit functions are given in the Supporting Information. Further, the use of equation (3) followed by pulse count at a different pump-power level enables the generation of BMCs with an optimised number of elementary constituents (see Supporting Information).

**2.2. Experimental results and discussion**

We start with the objective of obtaining a single-breather laser regime. To this end, we fix the pump power applied to the gain fibre to 70 mW. In a first series of experiments, starting from a unique, randomised set of polarisation parameters, we generate breathers without imposing any additional constraint on the features of the breather solution that is targeted. An example of an optimisation curve is presented in Fig. 2(a), which shows the evolution of the best and average merit scores of the population, as defined by equation (3), for each generation. We see that the best merit score quickly increases and converges to an optimised value after only 2 generations (ie., after 6 minutes). This fast convergence to the optimal state is facilitated by the relatively large population size of the generations. The average score of the population gradually increases to converge to almost the same value. The fluctuations of the average score, between the $6^{th}$ and $9^{th}$ generations, are ascribed, in some extent, to fluctuations of the measured features of established laser states, and predominantly to the mutation process. The spectral and temporal characteristics of the optimal state are summarised in panels (b-d) of Fig. 2. The various measurements confirm the operation of the laser in the targeted mode[5]: a RF spectrum exhibiting two symmetrical sidebands around the cavity repetition frequency (Fig. 2(b)), and a periodic compression and stretching of the optical spectrum over cavity roundtrips (Fig. 2(c)), accompanied by synchronous periodic changes of the pulse energy (Fig. 2(c), white curve), peak intensity and duration of the pulse in the time domain (Fig. 2(d)).

It is even more appealing to achieve intelligent control of the parameters of the breathers formed in the laser. By using our EA, we are able to generate breathers with tunable breathing ratio and period of oscillation. The breathing ratio is defined as the ratio of the largest to the narrowest width of the pulse spectrum within a period, $\Delta\lambda_{max}/\Delta\lambda_{min}$. Since the strength of the frequency sidebands in the RF spectrum is proportional to the breathing ratio, we can optimise the latter by taking into account the sidebands' strength in the definition of the merit function (see Supporting Information). Figure 3 shows the spectral and temporal dynamics of three examples of breathers with different breathing ratios that can be generated in the laser cavity by setting corresponding values in the merit function. They refer to the weakest breathing regime found in the cavity, which features a breathing ratio of 1.076 (Fig. 3(a)), the strongest breathing regime with a breathing ratio of 1.816 (Fig. 3(g)), and a moderate breathing regime (Fig. 3(d), breathing ratio of 1.471). By comparing the pulse spectra at the round-trip numbers of maximal and minimal spectrum extent within a period in panels (c, f, i) of Fig. 3, we can see that a larger breathing ratio is not achieved by an increased width of the widest spectrum but by a decrease of the narrowest spectrum extent. This is due to the fact that the maximum width of the spectrum is saturated by the gain bandwidth of the gain fibre. We have confirmed that 1.816 is the maximum breathing ratio achievable in the laser by manually tuning the pump power and the EPC under the single-breather mode-locked laser operation, with the results showing that the laser mode locking is destroyed when trying to increase the breathing ratio above 1.816.

Another important parameter of the breathing solitons that it is desirable to control is their period of oscillation. Indeed, it was recently found that the breathing period is a key parameter for the characterisation of subharmonic entrainment[4, 6]. Although this parameter is somehow related to breathing ratio, it is relevant to adapt the merit function to the specific optimisation of this feature (see Supporting Information), thus avoiding the need for an empirical manual adjustment of the gain/loss properties of the laser cavity. The results are summarised in Fig. 4, which shows the spectral and temporal dynamics of three examples of breathers with different oscillation periods. These results confirm that the designed merit function (cf. Supporting Information) can indeed be reliably used to tune the breathing period automatically. The maximum and minimum breathing periods found in the laser equal 251 cavity RTs (Fig. 4(a,b)) and 103 RTs (Fig. 4(e-f)), respectively.

In like manner to their stationary counterparts,[22, 23, 25, 47] breathing solitons in a laser cavity can interact and, within specific cavity parameter ranges, form robust multi-breather bound states,[5, 29] also termed breather molecules. Soliton molecules can be generated in fibre lasers by solely increasing the pump power above the fundamental mode locking regime, with the number of solitons within a molecule scaling with the pump power.[22, 25] However, the procedure for the excitation of breather molecules is not so straightforward. Further to this, it is difficult to generate multi (>2)-breather complexes in a normally dispersive fibre laser cavity because in the normal-dispersion propagation regime breathers do not emit dispersive waves.[5] It is therefore of great interest to implement an EA for the generation of BMCs. To this end, we apply an optimisation procedure that involves using the merit function given in equation (3) when the pump power is set to a level that favours multi-pulse self-starting of the laser, and subsequently applying pulse count (cf. Support Information) to control the number of breathers in the established multi-breather bound states. As an example, we show here the generation of BMCs composed of two, three and four elementary breathers. The optimisation time varies from a minimum of 2 min to a maximum of ~40 min.

We first perform an EA-based search for the laser polarisation parameters leading to the formation of breather-pair molecules ("diatomic" molecules). In these experiments, the pump power is increased to 120 mW. Considering that the individuals of a population are scored only against the number of breathers constituting the formed multi-breather states, several types of breather-pair molecules with very different dynamics can be accessed. Two such examples are presented in Fig. 5, which gives the round-trip evolutions of the DFT spectra, the first-order single-shot optical autocorrelation traces computed by Fourier transform of the DFT spectra, and the relative phases within the molecules retrieved from the autocorrelation traces.[25]

The recording of the real-time spectral interferogram for successive roundtrips shown in Fig. 5(a) displays a very dense pattern of spectral fringes (a magnified version is given in Fig. 5(b)). The extremely small spectral-fringe separation corresponds to a large pulse separation of 268 ps within the molecule (Fig. 5(c)). The relative phase $\phi_{21}$ between the trailing and leading breathers features an approximately linear evolution with time (number of cavity roundtrips) (Fig. 5(d), red curve). Since the slope of the phase evolution function is proportional to the intensity difference between the two bound pulses,[22, 23, 48] the quasi-linear phase evolution shown in Fig. 5(c) indicates an almost fixed intensity difference between the two synchronously evolving breathers, with the trailing breather being more intense than the leading one. A second type of a breather diatomic molecule found by the EA is presented in Fig. 5(e-h). In this case, we observe a significantly larger breathing of the optical spectrum, and an almost threefold reduction of the intra-molecular pulse separation (the pulse separation being 98 ps). In sharp contrast to the first molecule (Fig. 5(d)), the dynamics of the relative phase feature a pronounced oscillating behavior (Fig. 5(h), red). This indicates that the two breathers continuously exchange energy with each other. At the roundtrip numbers where the phase evolution function has extrema, the two breathers feature equal intensities, and the total energy is highest (Fig. 5(h), black).

Bound states of three breathers are also found in the laser using the EA when the pump power is increased to 150 mW. In a similar manner to the breather-pair case, the EA allows us to find different bound breather triplets, which are representative of three breather complex categories: (2+1) and (1+2) BMCs, and breather-triplet molecules. A (2+1) or (1+2) BMC originates from the stable binding of a breather-pair molecule and a single breather, with the breather-pair molecule being at the leading or trailing edge of the complex, while a breather-triplet molecule ('triatomic' molecule) comprises three nearly equally spaced breathers. The results are summarised in Fig. 6. In all breather complex cases, the DFT-based single-shot spectral measurements and the spatio-temporal intensity evolution clearly show a periodic breathing of the optical spectrum accompanied the synchronous variations of the pulse intensities. The pulse separations within a complex can be readily resolved from the spatio-temporal intensity map. For example, Fig. 6(d) shows that the intra-molecular separation in the leading breathing-pair molecule is 86 ps, while the trailing breather is 212-ps apart from the molecule. The spectral and temporal dynamics of the three breather complexes are markedly different from each other, and this also impacts the dynamics of the relative phases and the energy exchanges within the complexes (Supplementary Fig. 3).

Our EA can also be employed to search for bound states of four breathers in the laser under a pump power of 170 mW. Figure 7 depicts the dynamics of two examples of (1+3) BMCs, arising from the binding of a single breather and a breather-triplet molecule. The spatio-temporal intensity evolutions shown in panels (d) and (h) reveal very different pulse separations within the two complexes. Again, from more dedicated measurements, it is possible to gain insight into the BMCs' internal motion and appreciate the various behaviours of the different breather complexes, which differ from those usually exhibited by stationary soliton molecular complexes (Supplementary Fig.4). It is noteworthy that other types of quadri-breather complexes, such as (3+1) and (2+2) BMCs, were also found in the laser, but the characterisation of their internal dynamics was limited by the electronic-based spectrum resolution of our system, hence only partial information was obtained (Supplementary Figs. 5 and 6).

## 3. Conclusion

We have demonstrated, for the first time, the possibility of using EAs to perform search and optimisation of the breathing soliton regime in a fibre laser cavity. Through exploration of the nonlinear cavity dynamics, which can be accessed by automated control of the NPE transfer function, we have shown that composite merit functions, derived from specific features of the RF spectrum of the breather laser output, permit to achieve single-breather states with tailored parameters such as the breathing ratio and period of oscillation. At laser pump powers that favour multi-pulse emission, different types of BMCs with a controllable number of elementary constituents have also been generated automatically in the laser cavity. Our experimental setup combines a computer-interfaced single measurement device (an oscilloscope) with the quick setting of intracavity parameters through electronic polarisation control, thus allowing the EA to optimise the breather regime within a realistic time scale. While the EA concept presented here is ideally suited to mode-locked lasers, the breather-tailored merit functions designed in this work could also benefit the exploration of breather waves and related nonlinear dynamics in other systems, such as microresonators,[2, 3] fibre Kerr resonators,[1] and single-pass fibre systems.[9, 49] Contrary to the generation regimes of stationary pulses that have been mainly addressed by previous works using EAs, breathing solitons exhibit a fast evolutionary behaviour. In this respect, our work opens novel opportunities for the exploration of highly dynamic, non-stationary operating regimes of ultrafast lasers, such as soliton explosions, non-repetitive rare events and intermittent nonlinear regimes.[50]

The generation and propagation of pulses in multimode fibre systems have recently drawn great attention.[51-54] The nonlinear multimodal interference in multimode fibres has intensity discrimination properties that can be applied in mode-locked fibre lasers to generate a variety of different types of ultrashort pulses.[55] In these emerging laser designs, the vast parameter space makes systematic exploration impracticable, yet ideally suited to optimisation by an EA. Another promising area of research will be the expansion of the EA approach to use a wider range of tools in the general field of machine learning. Neural networks, for example, have previously been applied to the control of pulse shaping[56] and the classification of different regimes of nonlinear propagation[57,58] in single pass fibre geometries, and the optimisation of white-light continuum generation in bulk media.[59] Their extension to active control of mode locking has already been studied theoretically.[31] This extension appears a natural next step in the field, and with these techniques it may even be possible to control a broader range of processes within the plethora of complex nonlinear dynamics of mode-locked lasers.

**Supporting Information**

Text: 1, Details of the mutation.

2, Merit functions for:

breathing ratio tuning;

breathing period tuning;

breather molecular complex generation.

Fig. S1: Relaxation oscillations give rise to RF sidebands.

Fig. S2: Noise-like pulse emission also shows frequency sidebands in the RF spectrum.

Fig. S3: Internal phase dynamics of BMCs formed of three breathers.

Fig. S4: Internal phase dynamics of BMCs formed of four breathers.

Fig. S5: Dynamics of a (3+1) BMC.

Fig. S6: Dynamics of a (2+2) BMC.

**Acknowledgements** We acknowledge the support from National Key Research and Development Program (2018YFB0407100), the National Natural Science Fund of China (11621404, 11561121003, 11727812, 61775059, and 11704123), Key Project of Shanghai Education Commission (2017-01-07-00-05-E00021), Science and Technology Innovation Program of Basic Science Foundation of Shanghai (18JC1412000) and National Key Laboratory Foundation of China (6142411196307).

**Competing interests:** There are no financial competing interests.

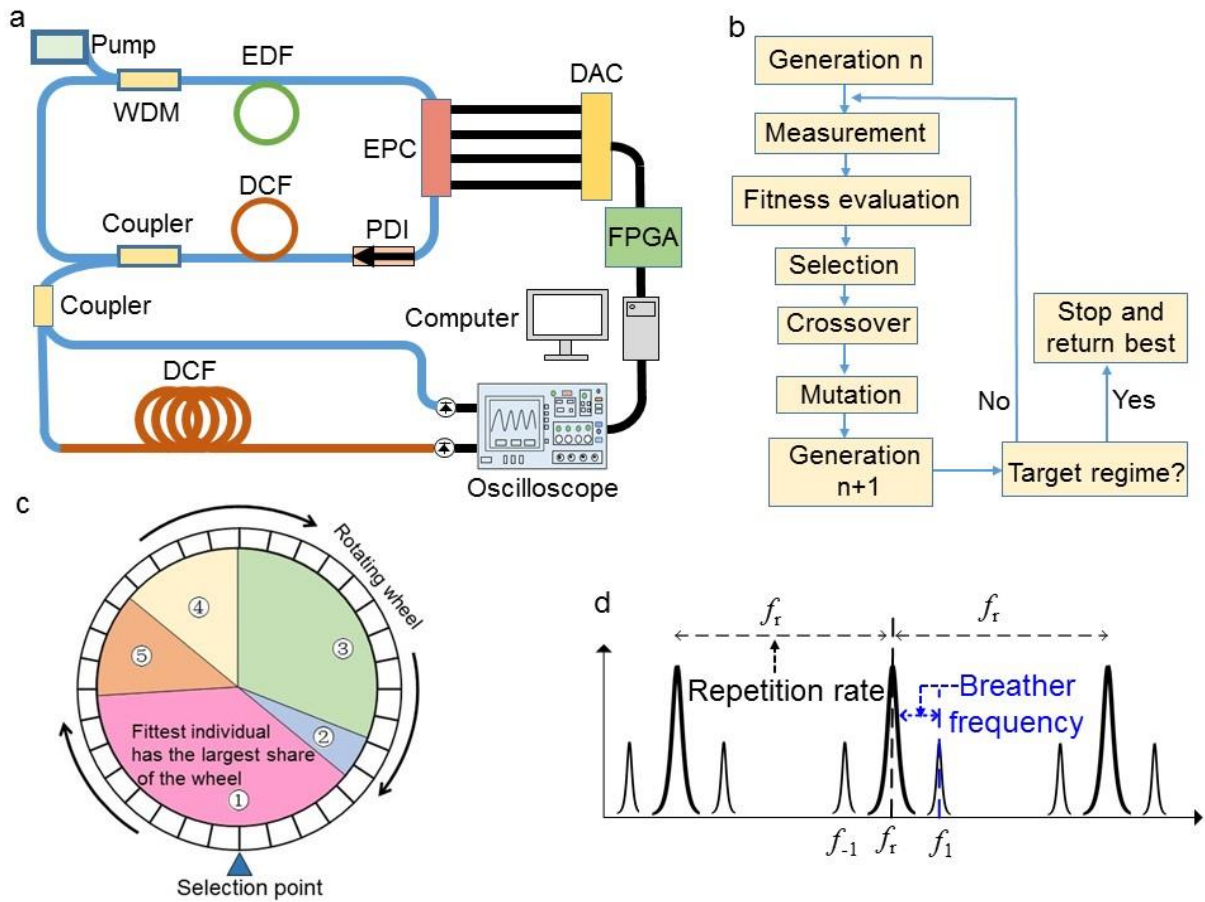

Fig. 1 (a) Experiment setup of the self-optimising breather mode-locked laser. WDM: wavelength-division multiplexer; EDF: erbium-doped fibre; EPC: electronic polarisation controller; PDI: polarisation dependent isolator; DCF: dispersion-compensating fibre; FPGA: field programmable gate array; DAC: digital-to-analogue converter. (b) Illustration of the EA principle. (c) Schematic of the 'roulette wheel' selection. (d) Sketch of the RF signal under breather mode locking, where $f_r$ is the cavity repetition frequency, and the sideband frequencies $f_{\pm 1}$ are a manifestation of breathers with the oscillation frequency $|f_{\pm 1} - f_r|$.

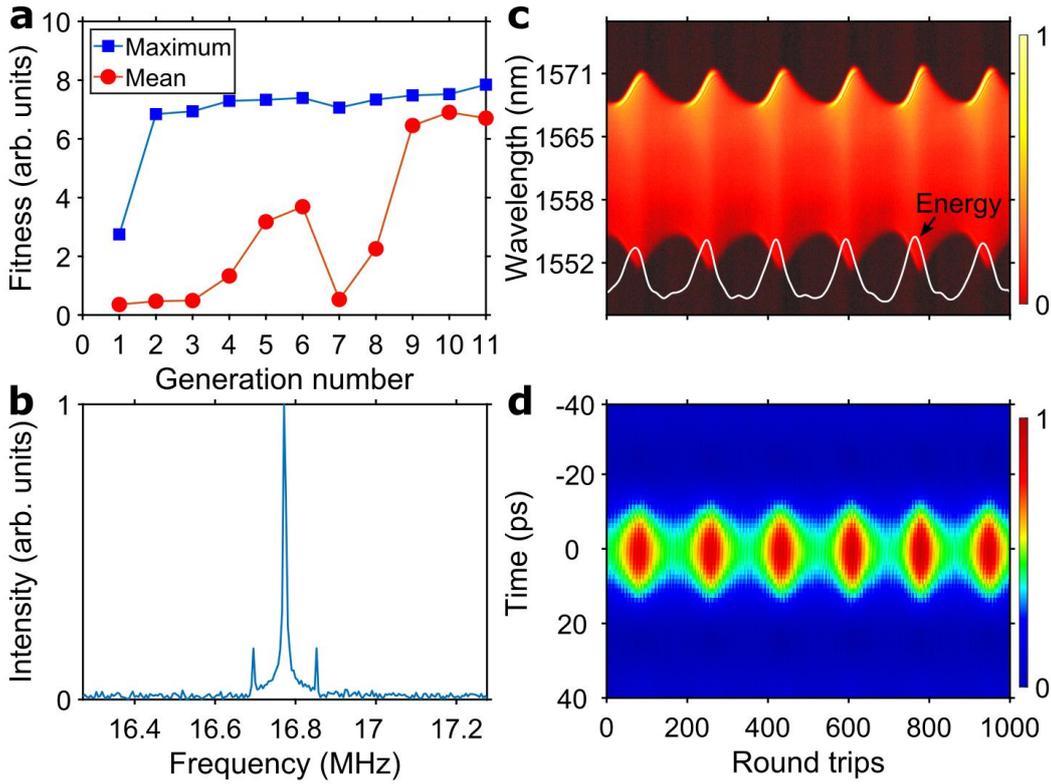

Fig. 2 (a) Evolution of the average (red circles) and maximum (blue squares) merit scores over successive generations, for the merit function given in Eq. (3). (b-d) Characteristics of the optimised state: (b) RF spectrum obtained by Fourier transform of the signal from the photodiode. (c) DFT recording of single-shot spectra over consecutive cavity round trips; the white curve represents the energy evolution. (d) Temporal evolution of the intensity relative to the average round-trip time over consecutive round trips.

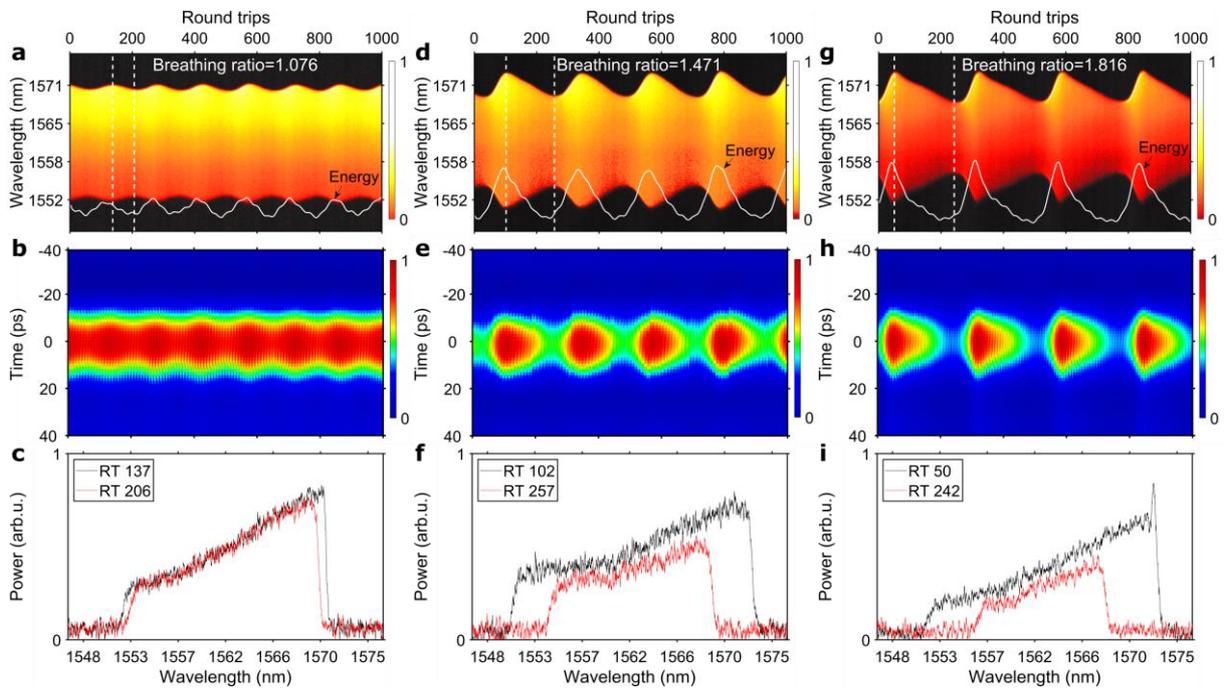

Fig. 3 EA optimisation results for breathing solitons with a tunable breathing ratio: dynamics of breathers with (a-c) small, (d-f) moderate and (g-i) large breathing ratios. (a,d,g): DFT recording of single-shot spectra over consecutive cavity round trips. (b,e,h): Temporal evolution of the intensity relative to the average round-trip time over consecutive round trips. (c,f,i): Single-shot spectra at the round-trip (RT) numbers of maximal and minimal spectrum extents within a period.

WILEY-VCH

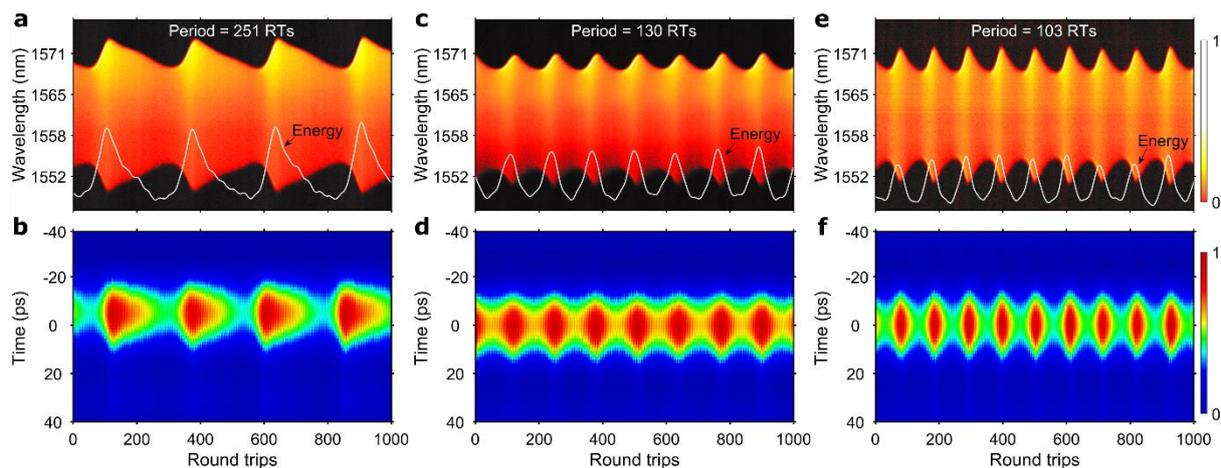

Fig. 4 EA optimisation results for breathing solitons with a tunable oscillation period: Dynamics of breathers with (a-b) large, (c-d) moderate and (e-f) small oscillation periods. (a,c,e): DFT recording of single-shot spectra over consecutive round trips. The white curve represents the energy evolution. (b,d,f): Temporal evolution of the intensity relative to the average round-trip time over consecutive round trips.

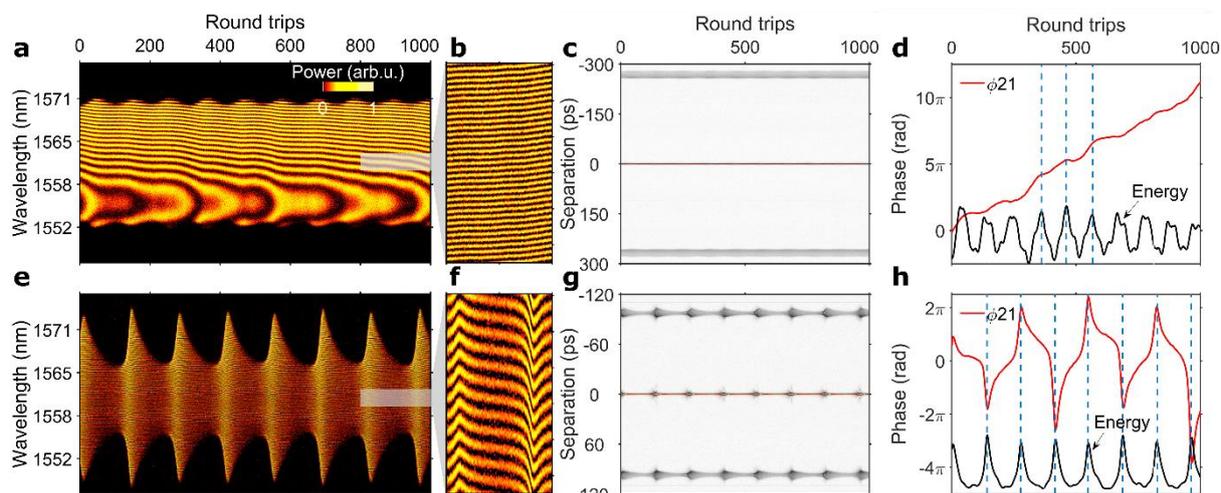

Fig. 5 Typical EA optimisation results for breather-pair molecules: Dynamics of (a-d) "increasing-phase" and (e-h) "oscillating-phase" breather molecules. (a,e) DFT recording of single-shot spectra over consecutive cavity round trips. A Moiré interference pattern is visible in panel (a). (b,f) Close-up view of the DFT recording. (c,g) Evolution of the first-order single-shot autocorrelation trace over consecutive round trips. (d,h) Evolution of the phase difference between the two breathers (red curve) and the energy of the molecule (black curve) as a function of the RT number.

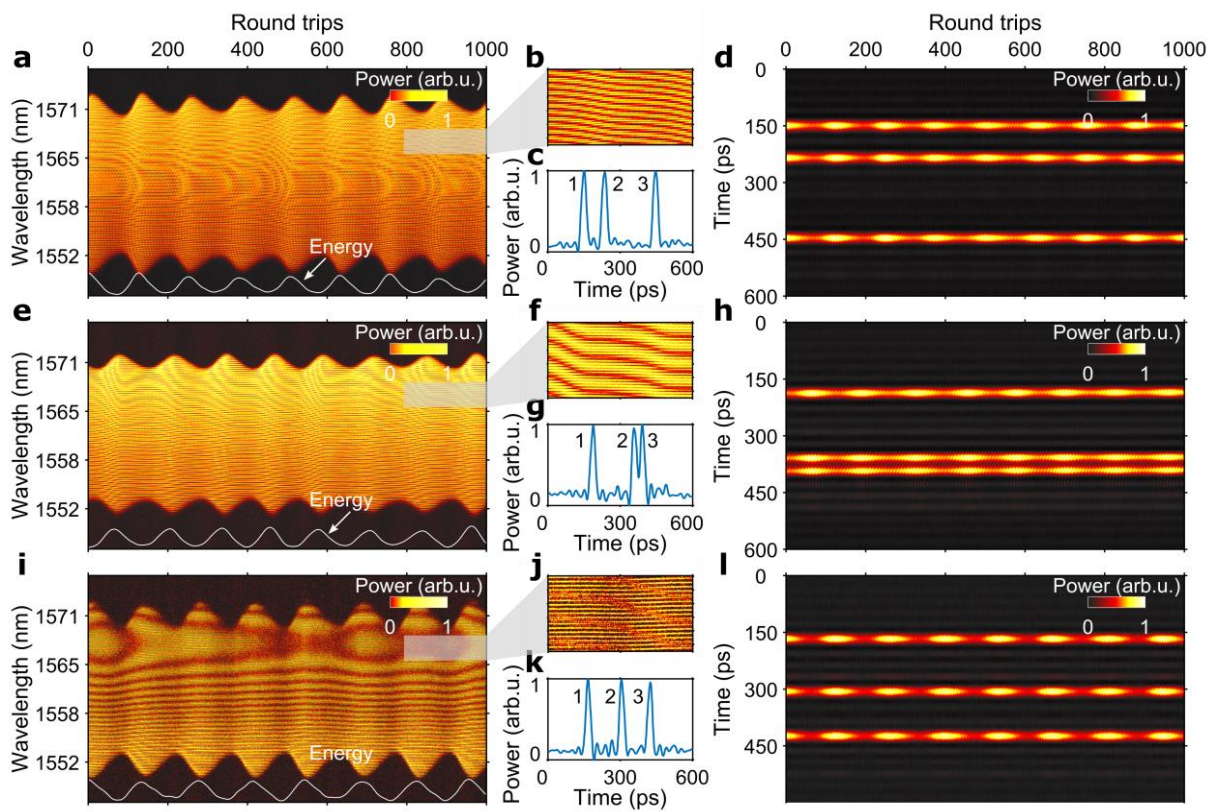

Fig. 6 Typical EA optimisation results for BMCs formed of three breathers: Dynamics of (a-d) a (2+1) BMC, (e-h) a (1+2) BMC, and (i-l) a breather-triplet molecule. (a,e,i) DFT recording of single-shot spectra over consecutive cavity round trips. The white curve represents the energy evolution. (b,f,j) Close-up view of the DFT recording. (c,g,k) Corresponding temporal intensity. (d,h,l) Temporal evolution of the intensity relative to the average round-trip time over consecutive round trips.

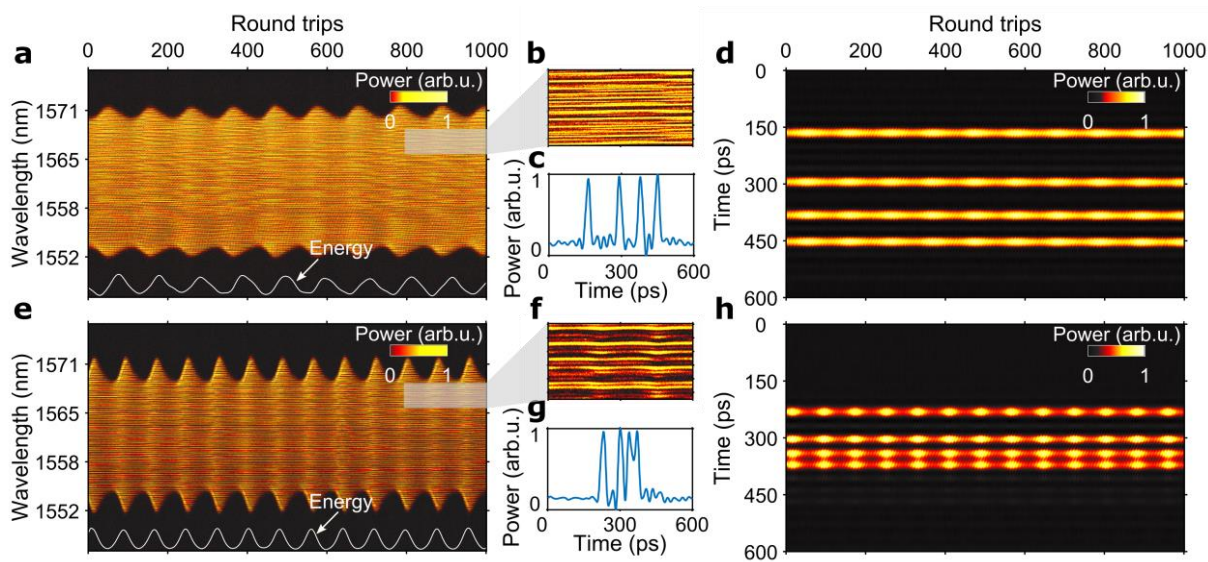

Fig. 7 Typical EA optimisation results for BMCs formed of four breathers: Dynamics of (1+3) BMCs with (a-d) large and (e-h) small internal pulse separations. (a,e) DFT recording of single-shot spectra over consecutive cavity round trips. The white curve represents the energy evolution. (b,f) Close-up view of the DFT recording. (c,g) Corresponding temporal intensity. (d,h) Temporal evolution of the intensity relative to the average round-trip time over consecutive round trips.

23